\documentclass{emulateapj}

\newcommand{\Rs}{$R_S$}
\newcommand{\rg}{$r$}
\newcommand{\R}{$R$}
\newcommand{\K}{$K$}
\newcommand{\Ks}{$K_S$}
\newcommand{\Klim}{$K_{lim}$}
\newcommand{\Kslim}{$K_{S, lim}$}
\newcommand{\IK}{$I$$-$$K$}
\newcommand{\RK}{$R$$-$$K$}
\newcommand{\RKs}{$R$$-$$K_S$}
\newcommand{\RsKs}{$R_S$$-$$K_S$}
\newcommand{\rgKs}{$r$$-$$K_S$}
\newcommand{\RKA}{$R$$-$$K_S>5.3$}
\newcommand{\RKB}{$R$$-$$K_S>6.0$}
\newcommand{\sext}{\textsc{SExtractor}}
\newcommand{\um}{$\mu$m} %
\newcommand{\Msun}{ $M_{\odot}$} %
\newcommand{\zs}{$z$$\sim$} %
\newcommand{\zform}{$z_f$} 
\newcommand{\Lstar}{$L^*$}
\newcommand{\Vaisanen}{V\"ais\"anen}

\slugcomment{accepted for publication in ApJ} 

\shorttitle{Overdensity of Extremely Red Objects}
\shortauthors{Sawicki et al.}

\begin{document}


\title{Properties of Extremely Red Objects in an Overdense Region
\altaffilmark{1}}

\author{
Marcin Sawicki\altaffilmark{2,3}, Matthew Stevenson\altaffilmark{2,4},
L.\ Felipe Barrientos\altaffilmark{5}, Brett Gladman\altaffilmark{6},
Gabriela~Mall\'en-Ornelas\altaffilmark{7}, and
Sidney~van~den~Bergh\altaffilmark{2} }

\altaffiltext{1}{Based on observations collected with the Very Large 
Telescope at the European Southern Observatory, Chile, as part of ESO programs
65.H-0543(A), 65.S-0513(A) and 69.A-0414(A).}

\altaffiltext{2}{
Dominion Astrophysical Observatory, 
Herzberg Institute of Astrophysics,
National Research Council, 
5071~West Saanich Road, 
Victoria, B.C., V9E 2E7, 
Canada
}
\altaffiltext{3}{
Present address: 
Department of Physics, 
University of California, 
Santa Barbara, 
CA 93106, USA
}

\altaffiltext{4}{
Department of Physics and Astronomy, 
University of Victoria, 
P.O. Box 3055, 
Victoria, B.C. V8W 3P6, 
Canada
}
\altaffiltext{5}{
Departamento de Astronom\'ia y Astrof\'isica,
Faculdad de F\'isica,
Pontificia Universidad Cat\'olica de Chile,
Casilla 306, Santiago,
Chile
}
\altaffiltext{6}{
Department of Physics and Astronomy,
University of British Columbia,
6224 Agricultural Road,
Vancouver, B.C., V6T 1Z1,
Canada
}
\altaffiltext{7}{
Harvard-Smithsonian Center for Astrophysics, 
Mail Stop 15, 
60 Garden Street, 
Cambridge, MA 02138,
USA
}

\begin{abstract}

We use a serendipitously discovered overdensity of extremely red
objects (EROs) to study the morphologies and cumulative surface number
density of EROs in a dense environment.  Our extremely deep imaging
allows us to select very faint EROs, reaching \Ks=21, or $\sim$2
magnitudes fainter than the \Lstar\ of passively evolving ellipticals
at $z$=1.5.  We find that the shape of the ERO cumulative surface
number density in our overdense field mimics that of the field ERO
population over all magnitudes down to \Ks=21 but with a factor of
3--4 higher normalization.  The excellent seeing in our images
(0.4\arcsec\ in \Ks\ and 0.6\arcsec\ in \R) allows for morphological
classification of the brighter (\Ks$<$19) EROs and we find a mix of
morphologies including interacting systems and disks; the fraction of
pure bulges (at most 38\%), galaxies with disks (at least 46\%), and
interacting systems (at least 21\%) is consistent with morphological
fractions in field ERO studies.  The similarity in the shape of the
cumulative surface density and morphological mix between our overdense
field and the field ERO population suggests that ERO galaxies in
overdense regions at $z$$\sim$1--2 may not have had an appreciably
different history than those in the field.

\end{abstract}


\section{INTRODUCTION}

Extragalactic extremely red objects (EROs) are galaxies that have
colors consistent with those expected of $z$$\gtrsim$1 passively
evolving old stellar populations.  As such, EROs are often used to
select what are hoped to be direct, passively evolving progenitors of
present-day ellipticals in the early universe.  However, the extreme
red color of EROs (e.g., \RKA, \RKB, $I$$-$$K>4$, etc.)  can also be
produced by heavily dust-obscured star-bursting galaxies.  Despite this
possible dichotomy, EROs continue to attract considerable interest
since either scenario can represent the direct progenitors of
present-day massive galaxies, albeit representing different formation
redshifts and histories.

Follow-up studies of ERO samples show that the ERO population is
indeed heterogeneous.  Spectroscopy reveals that the population is
made up of 50--70\% absorption-line objects and 30-50\% star-forming,
emission-line systems (Cimatti et al.\ 2002; Yan, Thompson, \& Soifer
2004a).  The distribution of EROs in $IJK$ or $RJK$ color-color space
also suggests a mix of starburst and quiescent systems (e.g., Pozzetti
\& Mannucci 2000; Smail et al.\ 2002; Gilbank et al.\ 2003; \Vaisanen\ 
\& Johansson 2004a; but see Moustakas et al.\ 2004) and Spitzer mid-IR 
observations show that $\sim$50\% of EROs indeed have high star
formation rates ($\gtrsim$12\Msun/yr, Yan et al 2004b; see also Smail
et al 2002 for VLA results).  X-ray observations suggest that a small
fraction (5--10\%) of EROs host active galactic nuclei (e.g., Roche,
Dunlop, \& Almaini 2003).  Imaging studies of EROs show a variety of
morphologies, including bulge-dominated objects, disk galaxies, and
interacting systems (Smith et al.\ 2002; Roche et al.\ 2002; Yan \&
Thompson 2003; Gilbank et al.\ 2003; Moustakas et al.\ 2004).  The
strong clustering of EROs (Daddi et al.\ 2000a, McCarthy et al.\ 2001)
suggests that at least part of the population is associated with
massive dark matter halos, and indeed absorption-line EROs are
strongly clustered (suggesting they are passively evolving
ellipticals) while emission-line ones cluster more weakly (Daddi et
al.\ 2002).  The picture of the ERO population is then one of a
heterogeneous mix consisting of both quiescent and star-forming
objects, some (but not necessarily all) of which may be the direct
progenitors of present-day massive elliptical galaxies.

EROs are a strongly clustered population and several very significant
ERO overdensities (possibly galaxy protoclusters?) have been reported
in the literature (e.g., Hall \& Green 1998; Cimatti et al.\ 2000;
Thompson, Aftreth, \& Soifer 2000; Best et al.\ 2003; Toft et al.\
2003; Wold et al.\ 2003; V\"ais\"anen \& Johansson 2004b).  All of
these overdensities have been found in dedicated searches that
targeted fields of high-$z$ AGN or IR-bright galaxies.  These searches
focused on confirming that high-$z$ AGN or IR galaxies are good
markers of large scale structures at high redshift and/or on studying
the spatial distribution of galaxies within them; very little work has
been done on systematically comparing ERO properties in overdense
regions with those of field EROs.  However, much can potentially be
learned from differences in galaxy properties as a function of
environment.  In the hierarchical galaxy formation paradigm, galaxies
are expected to have started forming first in overdense regions and so
EROs in overdensities may represent an older population than those in
the field.  At the same time, dense environments are thought to affect
galaxy {\it evolutionary} processes through mechanisms such as gas
stripping or mergers, resulting in different evolutionary histories in
field and cluster galaxy populations.  In either case, comparisons
between field EROs and EROs in overdense regions may teach us
important lessons about the process of galaxy formation in the early
universe.

In the present paper we present the results of a very deep study of
EROs in a serendipitously discovered overdense region.  The original
goal of our project was to constrain the surface number density of
very faint EROs by exploiting an existing extremely deep
($R_{lim}$$>$27) $R$-band image.  However, in the course of our
analysis we have discovered that we have chanced upon a strong but
rare ERO overdensity (see \S~\ref{surfacedensity}) and so have decided
to use our extremely deep and high-quality imaging to search for
environment-dependent differences in EROs.

Throughout this paper we assume $\Omega_M$=0.3, $\Omega_\Lambda$=0.7,
$H_0$=70 km~s$^{-1}$~Mpc$^{-1}$.  For consistency with other ERO work,
all magnitudes in this paper use the Vega normalization.  There are
several different definitions of the ERO color selection cut in the
literature (e.g., \RK$>$5.0, \RK$>$5.3, \RK$>$6.0, \IK$>$4.0) that
result in different redshift cuts and/or select for different star
formation histories.  As Yan \& Thompson (2003) point out, \IK\
selection likely favors systems with more prolonged star formation
than do \RK-selected surveys, and the redder cuts can be expected to
be more dominated by dust-enshrouded systems than the bluer cuts.  In
this paper we use the following two ERO definitions: \RKA\ and \RKB.

\section{OBSERVATIONS AND DATA PROCESSING}\label{data}

\subsection{Field Selection}\label{fieldselection}

Because of the extreme color of extremely red objects (\RKA\ or even
\RKB), ERO selection requires especially deep \R-band data and only 
relatively shallow infrared observations.  To study {\it very faint}
EROs --- to \Klim$\sim$21 --- we need extremely deep \R-band data
reaching \R$\sim$27 over a field large enough to contain a reasonably
large number of these fairly rare objects.  The acquisition of optical
data to such depths is very costly even with 8-meter class telescopes,
and so we decided to take advantage of existing ultra-deep \R-band
imaging that had already been obtained for another project.
Specifically, we used very deep VLT observations of a single pointing
that were originally taken to study Solar System objects (Gladman et
al.\ 2001) and that add up to 12 hours of integration on 8.2-meter
telescopes.  The availability of these extremely deep optical data
meant that we could be very economical by only needing to acquire
complementary infrared observations.

Our field, centered on R.A.=19:24:11, Decl.=$-$20:58:40 (J2000.0), is
located in the Ecliptic and close to the Galactic plane with Galactic
coordinates $l$=17:20:06, $b$=$-$16:21:18.  This location presents two
potential problems: that of foreground extinction and that of the
presence of a large number of Galactic stars.  We address these two
problems as follows.

Foreground extinction has the effect of dimming the fluxes and
reddening the colors of extragalactic objects. However, the amount of
extinction in our field, as measured from the Schlegel, Finkbeiner,
\& Davis (1998) dust maps, is small --- $A_R$=0.23 and $A_K$=0.03 --- 
with no evidence for significant variations in the vicinity of our
field.  Similarly small extinction values are found in the Burstein \&
Heiles (1982) reddening maps.  This amount of extinction does not
strongly affect the depth of our data and we explicitly correct for
its effects by adjusting our photometric zeropoints in
\S\S~\ref{Rimaging} and \ref{Kimaging}.

Very bright Galactic stars reduce slightly the effective area of our
survey because faint objects are more difficult to detect in the wings
of bright stars' light profiles and in the columns of saturated pixels
caused by spilled excess charge from saturated stars.  Additionally,
faint extremely red galactic stars provide a potential source of
contamination to our extragalactic ERO sample.  We account for these
effects by excluding areas affected by very bright stars from our
census (\S~\ref{surfacedensity}) and by constraining the stellar
contamination fraction in our catalog using morphological
characteristics (\S~\ref{stargalaxy}).  We note that the presence of
several bright stars in our field may eventually prove to be a
blessing as it allows the intriguing possibility of future follow-up
studies of ERO morphologies with adaptive optics.

\subsection{\R-Band Imaging}\label{Rimaging}

The $R$-band imaging was obtained on 2000 July 26-28 using the ESO VLT
8.2m Unit Telescopes ``Antu'' and ``Kueyen'' as part of a search for
faint Solar System objects (Gladman et al.\ 2001).  We used the FORS1
imaging spectrograph on UT1 and its FORS2 sibling on UT2.  Both FORS1
and FORS2 are equipped with 2k$\times$2k CCDs with 0.2 arcsecond
pixels, giving a field of view of 6.8\arcmin$\times$6.8\arcmin.

\begin{figure}
\plotone{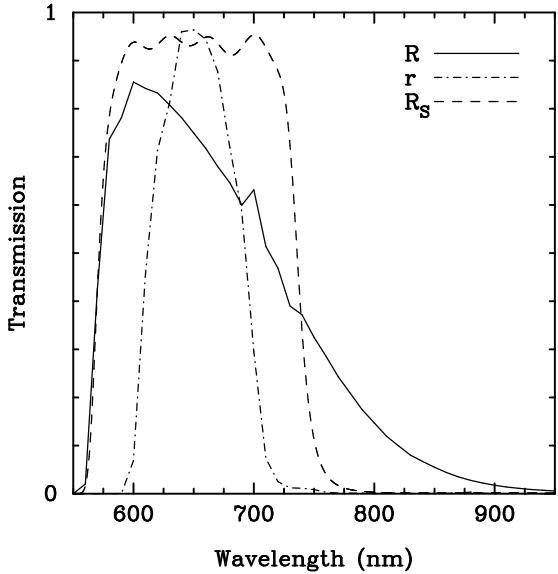}
\caption{
\label{filtercurves.fig} 
Filter transmission curves of the \R, \rg, and \Rs\ filters.  Our
observations were taken in \rg\ and \Rs, which have more tightly
defined bandpasses than the standard Bessel \R\ that is generally used
in ERO work.  }
\end{figure}

The FORS1 observations were taken in visitor mode and used the
Gunn-$r$ (hereafter, $r$) filter, while the FORS2 observations were
taken in service mode and used a wider ``Special $R$'' filter
(hereafter \Rs).  Both of these filters have a more sharply defined
bandpass than the standard Bessel $R$ filter that's commonly used in
ERO work (see Fig.~\ref{filtercurves.fig}).

All observations were taken in photometric weather and under
conditions of very good seeing.  The observations were broken into
short exposures, with the counts well within the linear regime of the
detector.  The $r$ observations (FORS1 in service mode) were dithered
between exposures, but the \Rs\ observations (FORS2 in visitor mode)
were undithered.  The total integration times were 35000s in $r$ and
9500s in \Rs, but the effective depths in the two filters were
comparable because the \Rs\ filter has a wider bandpass than the $r$
filter.

The individual frames were bias subtracted and flatfielded using
standard IRAF\footnote{IRAF is distributed by the National Optical
Astronomy Observatories, which are operated by the Association of
Universities for Research in Astronomy, Inc., under cooperative
agreement with the National Science Foundation} tasks before being
coadded into a final \Rs\ image and a final \rg\ image.

\begin{figure}
\plotone{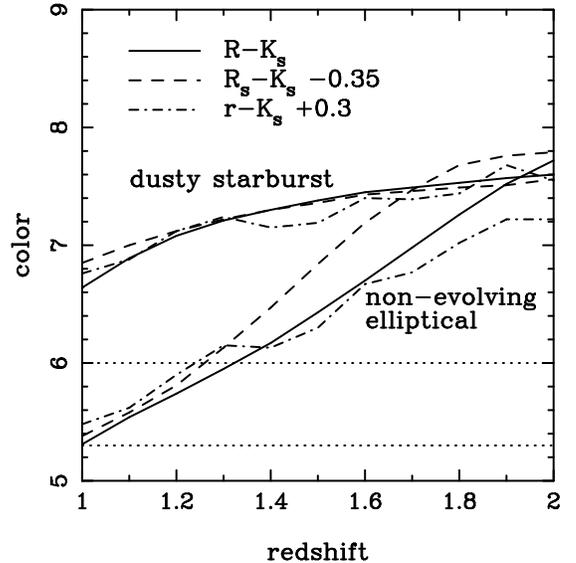}
\caption{ \label{modelcol.fig} Model colors of extremely red galaxies 
as a function of redshift.  The \RKs\ ERO selection thresholds of 6.0
and 5.3 are shown as dotted horizontal lines.  The solid, dashed, and
dash-dotted lines show the expected colors of dusty starbursts and
non-evolving ellipticals in three different photometric systems (\RKs,
\RsKs, \rgKs).  To compute these colors, we took the
50~Myr-old instantaneous burst from the Bruzual \& Charlot (1993)
spectral synthesis models substantially reddened with the Calzetti
(1997) dust law as well as the elliptical spectral energy distribution
from Coleman, Wu, \& Weedman (1980), and integrated both these SEDs
through the redshifted \R, \Rs, \rg, and \Ks\ filter transmission
curves.  The \RsKs\ and \rgKs\ tracks have been adjusted (by $-$0.35
and $+$0.3, as indicated in the figure) to bring them into agreement
with the \RKs\ models.  }
\end{figure}

The two images were then flux-calibrated as follows.  A first-order
flux calibration was applied to each of the two images based on
observations of standard stars. Next, the zeropoints were adjusted to
correct for foreground Galactic extinction as measured in the
Schlegel, Finkbeiner, \& Davis (1998) dust map ($A_R$=0.23, as
mentioned in \S~\ref{fieldselection}). Finally, the zeropoints were
further adjusted to compensate for the differences between our \Rs\
and \rg\ filters and the \R\ filter commonly used in ERO work. The
necessary adjustement was found by comparing model \Rs$-$\Ks,
\rg$-$\Ks, and \R$-$\Ks\ colors for several plausible ERO spectral 
energy distributions (SEDs) as shown in Fig.~\ref{modelcol.fig}. We
found that --- for EROs --- the \Rs\ zeropoint needed to be decreased
by 0.35 mag and the \rg\ one increased by 0.3 mag to bring them into
agreement with the \R-band scale.  We stress that this is a small {\it
relative} adjustment that transforms well the colors of extremely red
objects (but not necessarily those of normal, bluer galaxies) in our
\rg\ and \Rs\ images to the \RKs\ scale commonly used in ERO work.

After the adjustment of the zeropoints of the two master images, the
two images were inverse-variance weighted and combined.  The edges of
the resulting final image --- where the S/N was lower because of
dithering --- were then trimmed, giving a final area of 37.9
arcmin$^2$ with a total integration time of 44500 seconds.  The seeing
in the final image, measured on unsaturated stars, was 0.6\arcsec.

\subsection{\Ks-Band Imaging}\label{Kimaging}

The \Ks\ data were obtained in service mode using ISAAC on the 8.2m
VLT Unit Telescope 1 ``Antu'' during five nights between April 1, 2002
and May 8, 2002.  ISAAC's short-wavelength (1--2.5\um) camera is
equipped with a 1k$\times$1k Hg:Cd:Te array with 0.148 arcsecond
pixels, yielding an instantaneous field of view of
2.5\arcmin$\times$2.5\arcmin.  We covered the significantly larger
$R$-band image by dividing it into four quadrants for infrared
imaging.  In each quadrant we obtained 43 dithered \Ks\ exposures,
each consisting of the average of six 12s integrations, giving us a
total integration time of 3096s.

The individual exposures were flat-fielded, sky subtracted, and
coadded using the IRAF package DIMSUM.  Because DIMSUM did not
perfectly correct for the ISAAC bias (a known feature), the bias
residuals were subtracted in a separate, final step.  This residual
removal was done by simply calculating the median value for each row
of pixels and then subtracting the median from its respective row, as
suggested in the ISAAC Data Reduction Guide (Amico et al.\ 2002).

The photometric zeropoints were calculated separately for each night
using standard star images, and were corrected for foreground Galactic
extinction of $A_K$=0.03 as measured from the Schlegel, Finkbeiner, \&
Davis (1998) dust maps.  The individual images were then stacked
together into four master images, one for each quadrant.  Three of the
four stacked quadrant images comprised the full exposure depth of
3096s. Unfortunately, 22 of the 43 exposures in the north-eastern
quadrant had been taken on a night of particularly bad seeing and we
decided to exclude them from the final stacking, yielding a final
exposure time in that quadrant (only) of 1512s.

Finally, the four stacked quadrant images were astrometrically matched
to the $R$ band image and were combined into a single \Ks-band image
of the field.  This final \Ks-band image has a FWHM seeing of
0.4\arcsec, with virtually no variation from quadrant to quadrant.

\subsection{The Final Images}

\begin{figure}
\plotone{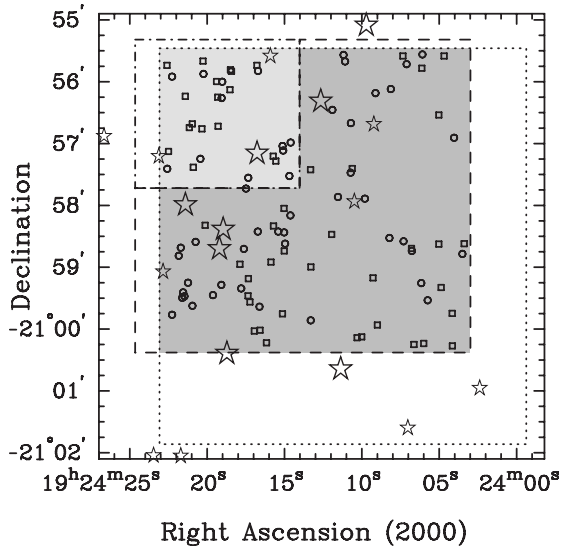}
\caption{\label{exptime.fig} Field geometry and the 
positions of EROs within the field.  The dotted line illustrates the
area covered by our combined \R\ image, the dashed line shows the
region of 3100s \Ks\ imaging, with the dot-dashed line box in the
north-east corner showing the shallower region of 1500s \Ks\ exposure.
The shaded area shows the area of our final, combined and trimmed
images: darker shading indicates the area that received the full 3100s
\Ks\ exposure and lighter shading shows the region that contains only
1500s of \Ks\ data.  Morphologically resolved EROs are shown as
squares while morphologically uncertain EROs are marked as circles.
Positions of bright stars from the USNO catalog are marked with star
symbols: the larger symbols show stars brighter than $R$=14 and the
smaller ones denote stars with 14$<$$R$$\leq$16.}
\end{figure}

The \R\ and \Ks\ images were trimmed to cover the region of mutual
overlap and a mask of bad pixels (introduced by the undithered \Rs\
data) and areas around bright foreground stars was created.
Figure~\ref{exptime.fig} shows the geometry of the final images.  The
size of the final image is 4.7$\arcmin$$\times$4.9$\arcmin$, which at
$z$=1.5 corresponds to 2.4~Mpc$\times$2.5~Mpc in the $\Omega_M$=0.3,
$\Omega_{\Lambda}$ = 0.7, $H_0$=70 km~s$^{-1}$~Mpc$^{-1}$ cosmology.
The final good area --- the area of the trimmed images that does not
contain bright foreground stars or bad pixels --- is 19.6~arcmin$^2$,
of which 15.9~arcmin$^2$ had been imaged to the full 3096s depth in
\Ks-band, and the remaining 3.7 arcmin$^2$ had the shallower exposure
time of 1512s.    As determined from simulations (see
\S~\ref{completeness}), the limiting magnitude of the \Ks\ image is
\Kslim=21.0 except for the shallower north-east corner, which reaches
\Kslim=20.6.

\section{THE ERO CATALOG}

\subsection{Source Detection and Photometry}\label{photometry}

We used the SExtractor package (Bertin \& Arnouts 1996) for object
detection and photometry.  Object detection was done on the \Ks-band
image, and we required 5 contiguous pixels to be above the threshold
of 2.2$\sigma_{sky}$ for a candidate to be considered an object.
\Ks-band magnitudes were measured within SExtractor's ``best'' 
apertures, which are Kron-like apertures for the majority of objects
except for objects suspected of having close companions, for which
isophotal magnitudes are used instead (Kron 1980; Bertin \& Arnouts
1996).  \RKs\ colors were measured in matched 1.0\arcsec-diameter
apertures on the \R-band image and on a copy of the \Ks-band image
that had been Gaussian-smoothed to match the poorer, 0.6\arcsec\
seeing of the \R-band data.  We checked the \Ks-band number counts of
objects classified as resolved by SExtractor and found that they are
in good agreement with galaxy counts in the literature (e.g.,
Bershady, Lowenthal, \& Koo 1998; Totani et al.\ 2001).

\subsection{Star-Galaxy Classification}\label{stargalaxy}

L- and M-type dwarfs are sufficiently red to be included in our ERO
color cut and so at the low Galactic latitude of our field our ERO
sample may be contaminated by a significant number of Galactic stars.
We used the morphological appearance of our objects in the 0.4\arcsec\
\Ks-band image to constrain the amount of such foreground stellar
contamination.  We classified the morphologies of our EROs both by
using automated neural-network classification in SExtractor and by
visual inspection by two of the authors (MS and MS). The human
classifiers categorized objects as either ``resolved'' or
``uncertain'', where ``resolved'' were those EROs which had clearly
extended morphologies that made them extremely unlikely to be
foreground stars.  The automated classification by SExtractor yielded
results consistent with the human classification. We present results
of the human star/galaxy ERO classification in the rest of this paper,
although we note that the results do not differ significantly if we
use the SExtractor neural network results.  Column 6 in
Table~1 lists the results of our Resolved/Uncertain
classification.

Throughout the rest of the paper our approach is to assume that EROs
classified as ``R'' (Resolved galaxies) are indeed high-$z$ galaxies,
whereas objects classified as ``U'' (Uncertain) may contain a mixture
of more compact galaxies and foreground stars.  The R subsample thus
gives a robust and very conservative {\it lower limit} on
extragalactic ERO number density, while the entire R+U sample gives a
conservative upper limit.

The ``R'' sample thus contains extragalactic EROs only, with no
stellar contamination.  However, the ``U'' sample should contain very
few stellar contaminants for the following reason.  At brighter
magnitudes (15$<$\Ks$<$19), our ERO catalog contains only 3/24 (13\%)
potential stars (see also Fig.~\ref{completecounts.fig}).  Since the
ratio of stars to galaxies generally decreases towards fainter
magnitudes, the stellar fraction is likely to be even lower at
\Ks$>$19 than it is at \Ks$<$19.  Most of the faint ``U'' objects are 
thus likely to be classified as such not because they are stars but
because they have low S/N that prevents us from seeing extended
structure.  Thus, while the U sample gives a very hard lower limit, we
can reasonably assume that the true counts of extragalactic EROs are
very close to those given by the combined R+U sample.

\subsection{Completeness}\label{completeness}

We used simulations to test the accuracy of our photometry and to
understand and correct for the incompleteness of our survey.
Specifically, we inserted artificial objects of known flux and color
at random positions in our images and then repeated the object-finding
and photometry procedures of \S~\ref{photometry} to determine the
completeness of our catalog and its photometric accuracy by measuring
the recovered fraction, magnitudes, and colors of the artificial
objects.  We surveyed the parameter space spanning a range of
magnitudes (17$\leq$\Ks$\leq$22) and colors (4$\leq$\RKs$\leq$8) with
a sufficiently large number of realizations (1000 artificial objects
at each seeing, color, and magnitude step) so as not to be dominated
by small number statistics.  We conducted these tests using artificial
objects with Gaussian profiles of 0.5\arcsec, 0.75\arcsec, and
1.0\arcsec\ FWHM.

\begin{figure}
\plotone{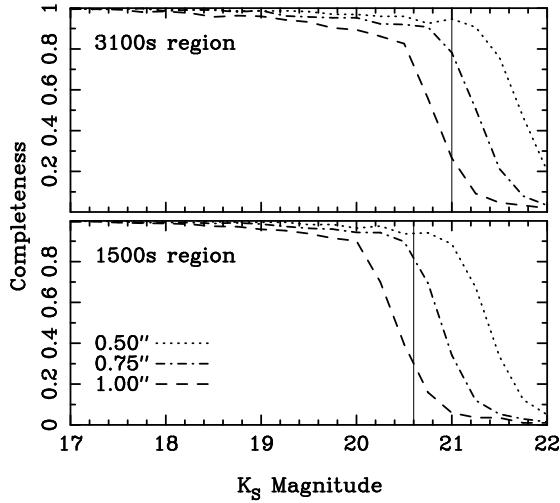}
\caption{\label{complete.fig} Detection completeness in the \Ks\ image 
derived using simulations described in \S~\ref{completeness}.  The top
panel illustrates the completeness in the 3100s integration region,
while the lower panel shows the completeness in the 1500s quadrant.
The three different curves are for objects with FWHM of 0.50\arcsec,
0.75\arcsec, and 1.00\arcsec.  The 0.75\arcsec\ curve was used to
correct for incompleteness of the ERO surface number density.
(\S~\ref{surfacedensity}).  The vertical lines indicate our adopted
80\% completeness limits. }
\end{figure}

Figure~\ref{complete.fig} shows the completeness functions determined
from our simulations.  As expected, completeness is higher at a given
magnitude in the deeper, 3100s portion of the image than in the
shallower, 1600s quadrant.  Also, not surprisingly, completeness
depends on the size of the object. Because the typical sizes of
objects in our ERO catalog (\S~\ref{photometry}) were
FWHM$\leq$0.75\arcsec, we decided to adopt the curves for 0.75\arcsec\
as a conservative incompleteness estimate. In the deeper, 3100s
portion of the image, the 80\% completeness is reached at
\Kslim$\sim$21, and in the shallower, 1600s quadrant, it is reached at
\Kslim$\sim$20.6.  We note that these limiting magnitudes are, at
80\%, conservative.  We could have defined fainter limits by choosing
50\% or even 20\% completeness but we chose 80\% completeness to
ensure results that are conservatively robust.

Our simulations also showed that there were no significant biases in
the measurement of colors of EROs down to our adopted completeness
limit, and that the brightness and color error estimates given by
SExtractor were accurate for our purposes.

\subsection{The ERO Catalog}

\begin{figure}
\plotone{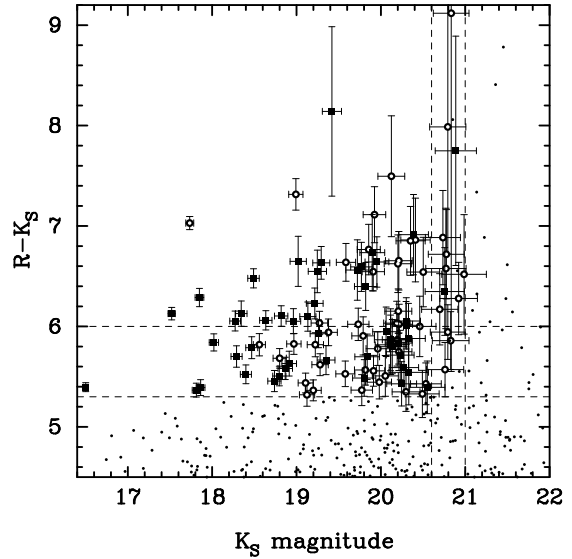} 
\caption{\label{colmag.fig} Color-magnitude diagram of red objects in 
our field.  Filled squares show objects morphologically classified as
resolved galaxies and open circles denote morphologically uncertain
objects. Non-EROs and objects fainter than the completeness limit are
plotted as dots.  The vertical dashed lines denote our completeness
limits and the horizontal ones indicate the ERO color selection cuts. 
Error bars show 1$\sigma$ uncertainties measured by
\sext.}
\end{figure}

EROs were identified solely on the basis of their \RKs\ colors.
Figure~\ref{colmag.fig} shows the color-magnitude diagram of all the
red objects in our field and Table~1 presents the
positions, \Ks\ magnitudes, and \R$-$\Ks\ colors of all EROs.  For the
purposes of our catalog, EROs are defined to be all objects that are
above our adopted completeness limits and are redder than
\RKs=5.3. They are plotted with large symbols in Fig.~\ref{colmag.fig}:
filled squares denote objects that are morphologically resolved and
therefore very likely distant galaxies, while open circles show
objects with uncertain morphologies. In total, to \Ks=21.0, our
catalog contains 101 \RKA\ of which 49 are also \RKB\ EROs.  It is
worth pointing out that our \R-band data are deep enough that all
\Ks-selected objects are also detected in \R.

\section{ANALYSIS}\label{analysis}

\subsection{Surface Number Density}\label{surfacedensity}

\begin{figure}
\plotone{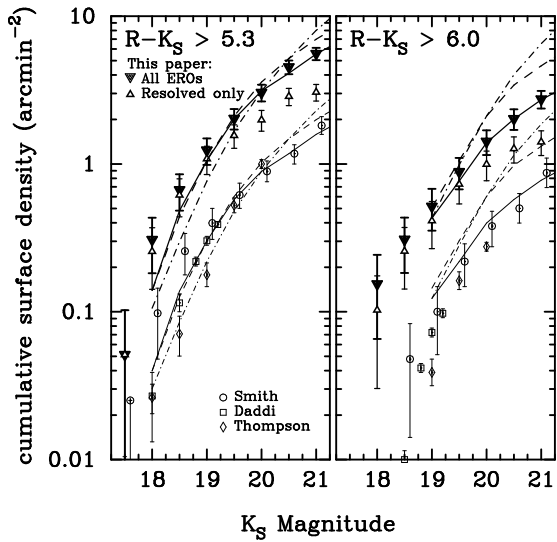}
\caption{\label{completecounts.fig} Cumulative surface number density 
of EROs for both color selection cuts.  Triangles show our counts, for
{\it all} EROs (downward pointing filled triangles) and for only those
thus bracket the actual surface number density of extremely red
galaxies in our field, although the true ERO counts are likely to be
close to the full counts (filled, downward pointing triangles), as we
argue in \S~\ref{stargalaxy}.  The data have been corrected for
incompleteness, although this correction is small.  Open symbols show
measurements from field ERO surveys (Thompson et al.\ 1999; Daddi et
al.\ 2000b; Smith et al.\ 2002).  The {\it thin} solid, dashed, and
dot-dashed curves are Daddi et al.\ (2000b) PLE models with
$\tau=0.1$Gyr and $z_f$=2.5, 3.0, and 10.0 (see Smith et al.\ 2002);
their vertical scaling is dictated by the observed luminosity function
of ellipticals in the local universe.  The upper, {\it thick} curves
are the very same PLE models simply shifted upwards in surface number
density by a factor of 3.5.}
\end{figure}

Figure~\ref{completecounts.fig} shows the cumulative ERO surface
number density in our field for both \RKA\ and \RKB\ EROs.  These
measurements have been corrected for incompleteness, although --- as
Fig.~\ref{complete.fig} shows --- the correction is small even at the
limit of our catalog.  Our surface number densities are shown both for
morphologically resolved EROs (upward-pointing triangles) and for all
EROs, including resolved and unresolved ones (downward-pointing filled
triangles).  As we pointed out in \S~\ref{stargalaxy}, these two
categories of EROs represent the extremes of the plausible
extragalactic ERO surface number density in our field in that the
former class is virtually guaranteed to be free of stellar
contamination thus giving a robust lower limit, whereas the latter
includes all possible extragalactic EROs --- including compact
galaxies --- and so gives an upper limit.  The true extragalactic ERO
surface number density thus lies somewhere between the bounds defined
by the upward- and downward-pointing filled triangles, although we
expect their true surface number density to be far closer to the upper
limit as explained in \S~\ref{stargalaxy}.

The most striking feature of Fig.~\ref{completecounts.fig} is that the
ERO surface number density in our field is $\sim$3--4 higher than
that seen in other surveys (e.g., Thompson et al.\ 1999; Daddi et
al.\ 2000b; Smith et al.\ 2002).  This is true for both the \RKA\ and
\RKB\ EROs, and holds at all but the faintest magnitudes, \Ks$\sim$21,
where the number of objects for which we are unable to make a
star/galaxy determination becomes large.  Even at this faint end,
however, our field is overdense in resolved EROs compared to other
surveys, and it is very likely that the overdensity remains as high as
a factor of $\sim$3--4 at all magnitudes \Ks$\leq$21 given the upper
limits on the ERO surface number density and the fact that --- as we
discussed in \S~\ref{stargalaxy} --- the true counts are actually
likely to be close to this upper limit.

Finally, we note the significant gradient in the surface number
density of EROs in our field.  As Fig.~\ref{exptime.fig} shows, the
ERO density increases from west to east, with the eastern half of the
field containing nearly twice as many EROs (65) as the western half
(36).  The excess in the eastern half may in fact be even higher than
that given that the north-eastern quadrant of our catalog has a
shallower completeness limit than the rest of the field.  In any case,
the gradient is highly significant, since a 65:36 or 36:65 ratio
occurs only $\sim$0.5\% of the time in a random distribution.  The
number density gradient raises in the direction of the 4~mJy 1.4GHz
radio source NVSS~J192430-205733 (Condon et al.\ 1998), which is
located only $\sim$2\arcmin\ from the eastern edge of our field (if we
assume that the 65:36 gradient has to rise {\it towards} the radio
source, then the probability of the gradient being a random artefact
is even lower, namely $\sim$0.25\%).  While no redshift or photometry
at other wavelength is available for this source, it is not
implausible that the source may reside in the ERO structure with the
ERO density gradient pointing towards it.

\subsection{A cluster of EROs?}

The most plausible explanation of this ERO overabundance is that we
have discovered a true physical overdensity of EROs.  The existence of
an ERO overdensity is not surprising given that EROs are known to be
strongly clustered (e.g., Daddi et al.\ 2000a; Roche et al.\ 2002),
although a factor of 3--4 excess is very rare in a random field the
size of ours: Daddi et al.\ (2000a) show that a $>$3-fold excess would
occur in fewer than 0.1\% randomly placed 25~arcmin$^2$ fields even
for luminous ($Ks$$\leq$18.8), possibly {\it more} clustered, EROs.
However, ERO overdensities as rich as the one reported here have been
discovered by surveys that target the fields of high-$z$ AGN (e.g.,
Hall \& Green 1998; Cimatti et al.\ 2000; Hall et al.\ 2001; Wold et
al.\ 2003), and mid-IR bright galaxies (V\"ais\"anen \& Johansson,
2004b).  For example, Wold et al.\ (2003) find that to their
\Ks$\sim$19 limit, the average number density of EROs in their 13 QSO
fields is a factor of 2--3 higher than that in random-field surveys,
and V\"ais\"anen \& Johansson (2004b) find an excess of 2--5 in a
number of small fields centered on ISO mid-IR selected objects.  At
any rate, our 3--4-fold density excess is entirely consistent with the
results of such surveys that target fields around known high-$z$
objects, although our imaging traces the excess to a significantly
fainter level (\Ks=21.0), or $\sim$2 magnitudes below
\Lstar at \zs1.5.

Our 3--4-fold surface overdensity may imply a much higher {\it
physical} overdensity.  Whereas the field ERO population is
distributed over a likely redshift range $z$$\sim$1--2, our excess
EROs may well be located in a protocluster or a filament, spanning a
volume that corresponds to a much smaller $\Delta$$z$.  We have no way
of knowing what that volume is with just the data in hand, but we can
make a very rough, illustrative estimate of the physical overdensity
as follows.  If we assume that the field ERO population uniformly
spans $z$=1--2 but that the EROs in our overdensity reside in a cube
with line-of-sight depth equal to the angular size of our field (i.e.,
$\sim$2.5Mpc$\times$2.5Mpc$\times$2.5Mpc, comoving units), then ---
after statistically correcting for foreground/background ERO
contamination based on field ERO counts --- the spatial density of
EROs in the structure is $\sim$5000 higher than that of the field ERO
population.  The factor of $\sim$5000 is a rough {\it upper limit} on
the overabundance of EROs in the structure.  If, for example, we
assume that the line-of-sight extent of the structure is 25~Mpc
instead of 2.5~Mpc then the spatial density in the structure is
reduced to $\sim$500 times that of the field population.  Similarly,
if the redshift depth occupied by the {\it field} ERO population is
smaller than the $\Delta z$=1 we assumed here, then the overdensity
factor will also be reduced.  In any case, however, as long as the
overdensity we have discovered is physical then the relatively small
3--4-fold number excess translates into a substantial spatial
overdensity that may rival that of rich clusters in the present-day
universe.

Finally, we note that it is possible that the radio source
NVSS~J192430-205733 located just off the eastern edge of our field may
be a high-$z$ object that is part of the same structure as the EROs.
This possibility is supported by the ERO density which increases in
the direction of the radio source.  However, at present no redshift
(or any other) information is available on this object and
confirmation of this hypothesis will need further observations.

\subsection{Comparing ERO populations in the overdensity and in the field}

The depth of our data and the excellent seeing in our images allows us
to study the properties of the ERO population in our overdense region.
Differences between the ERO population in this overdensity and in the
field may give interesting clues to the nature of EROs, to the origin
and evolution of present-day early-type galaxies, and to the
populating of large galaxy structures in the early Universe.  We focus
here on comparing the surface number density to \Ks=21 and on
morphologies of brighter, \Ks$<$19, EROs.

\subsubsection{Surface number density}\label{numbercountscomparison}

The ERO surface number density in our overdense field is higher than
in field ERO surveys at {\it all} magnitudes down to \K=21.  If --- as
we have argued in \S~\ref{stargalaxy} is likely the case --- our true
ERO surface number density is close to our upper limits
(downward-pointing filled triangles) in Fig.~\ref{completecounts.fig},
then the difference between our ERO overdensity and EROs in the
general field is consistent with a straightforward overall 3--4-fold
increase in number density, {\it irrespective of magnitude}.  This
similarity extends so far as to the presence of a break in the slope
of field ERO surface number density at \Ks$\sim$19--20 (Smith et al
2002; Roche, Dunlop \& Almaini 2003).  We conclude that the surface
number density in our ERO overdensity is consistent with a
straightforward increase in number density of EROs, irrespective of
ERO brightness.

The fact that the surface number density slope does not vary with
environment is surprising: we might have expected to see environmental
variations given that the ``galaxy formation clock'' is expected to
turn on earlier in overdense regions and that environment can modify
subsequent galaxy evolution in high-density regions through processes
such as tidal stripping, ram-pressure stripping, suppression of
galaxy-galaxy mergers, and the aging of stellar populations.

Daddi et al.\ (2002b), among others, have generated ERO surface number
density predictions using pure luminosity evolution (PLE) models which
attempt to model EROs as elliptical galaxies by assuming that they
form all of their stars at some high redshift, $z_f$, that these
stellar populations evolve passively ever since, and that at the epoch
we observe them they are already located in galaxies that follow a
luminosity function that by $z$=0 will fade to be consistent with that
of present-day ellipticals.  These PLE models can reproduce reasonably
well the field ERO surface number density, as is illustrated by the
thin curves in Fig.~\ref{completecounts.fig} (Smith et al.\ 2002).
The faint-end slope of these models compared with the field ERO
surface number density suggests a redshift of formation of
$z_f$$\sim$2.5 for the field \RKB\ EROs (the case for \RKA\ EROs is
less unambiguous but also consistent with this moderate formation
redshift).

Adjusting the Daddi et al.\ PLE models upwards in density by a factor
of 3.5 brings them into agreement with the surface number density of
{\it bright} EROs (both \RKA\ and \RKB) in our field (see
Fig.~\ref{completecounts.fig}).  The \zform=2.5 models are then in
good agreement with our ERO counts both for \RKA\ and \RKB\ EROs down
to our limit of \Ks=21.  This is again unexpected because we would
have expected galaxy formation to begin earlier in our overdense
region than in the field, with \zform$>$2.5 and hence a steeper
faint-end slope of the \RKB\ counts.  However, the models with higher
formation redshifts --- \zform=10 and even \zform=3 --- are steeper
than the observed counts at the faint end.

The fact that the same \zform=2.5 model (modulo a density
normalization) fits best both the field ERO population {\it and} the
ERO population in our overdensity suggests that field and cluster EROs
may share the same evolutionary history regardless of their present
environment.  Since neither population is well fit by the \zform=3
model further suggests that the field and cluster ERO populations may
in fact be co-eval or nearly co-eval.  We note that this argument is
not strongly sensitive to the details of the models: since we are
comparing two identically-selected populations that are best fit by
the same model, any systematic change in the model that may affect the
model's details (such as the \zform) will apply equally to the field
and cluster population.  Consequently, the conclusion about the
co-evalness of the cluster and field populations is fairly robust
under the assumption that the two populations are composed of similar
objects.  This conclusion can, however, be affected if the two
populations are composed of distinct subpopulations (e.g., a mix of
old, quiescent galaxies and dusty starbursts) that are present in
proportions that vary between the field and the cluster, but that
conspire to reproduce the observed magnitude dependence of the
cumulative surface number density.  We constrain this possibility of a
different mix of subpopulations next.

\subsubsection{Morphologies}\label{morphologies}

\begin{figure}
\plotone{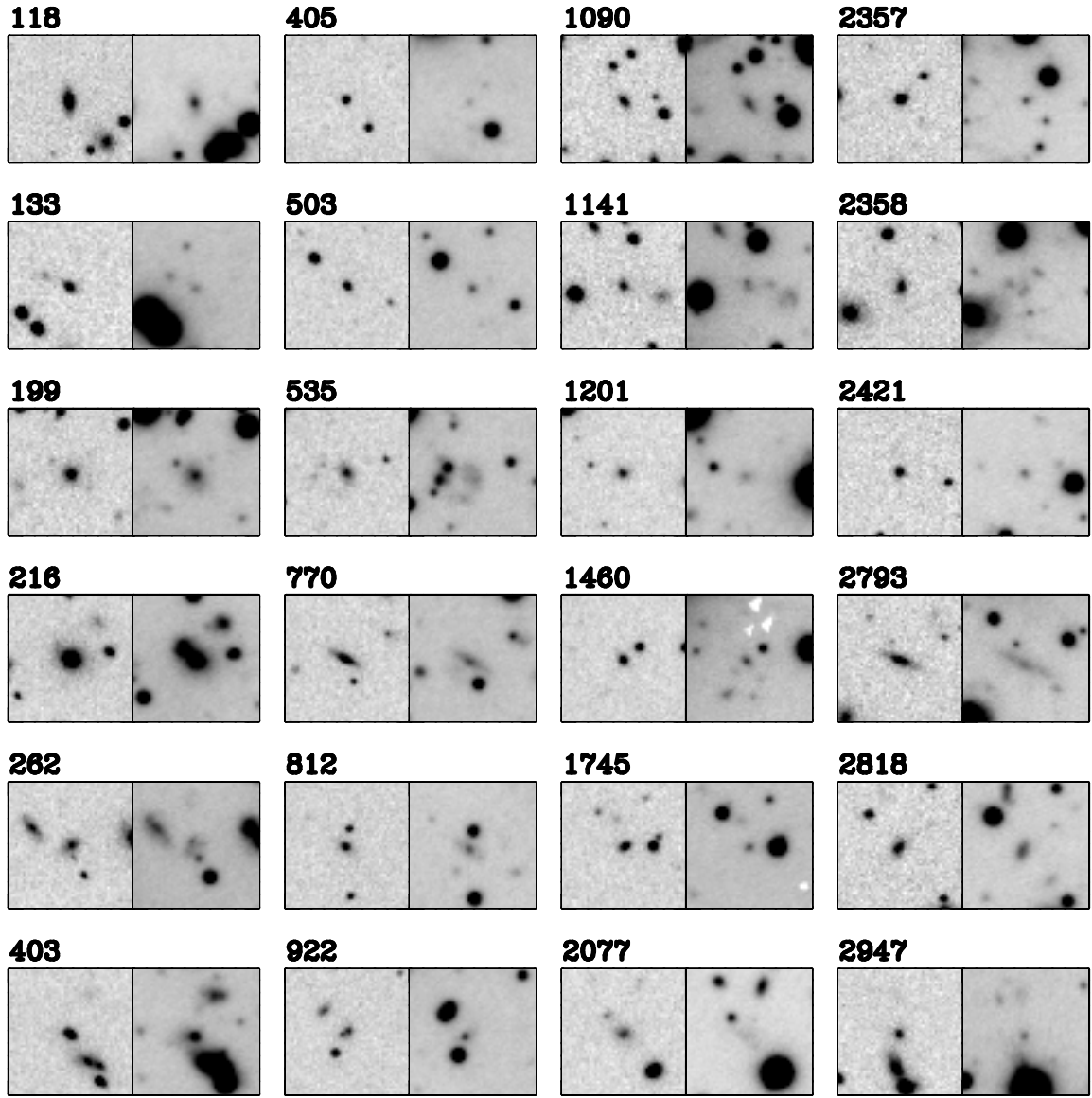}
\caption{\label{morph.fig} EROs with \Ks$<$19 and \RKA.  Shown are 
\Ks\ (left) and \R\ (right) image pairs. Each image is 
10\arcsec\ on the side and the ERO is at its center.  Seeing is
0.4\arcsec\ and 0.6\arcsec\ in the
\Ks\ and \R\ images, respectively.}
\end{figure}

At low and intermediate redshifts the morphological mix of galaxies
depends on environment, with denser environments being significantly
richer in early-type galaxies than the field (e.g., Hubble 1936;
Dressler 1980; Hashimoto \& Oemler 1999; Goto et al.\ 2003). If the
same environmental trends hold at high redshift, we should expect more
passively evolving old stellar populations and fewer dusty starburst
EROs in our overdensity than in the field.  Given the excellent
0.4\arcsec\ seeing in our \Ks\ image, we decided to compare the
morphologies of the brighter EROs in our sample with those of field
ERO morphological studies in the literaure.  We focus on EROs brighter
than \Ks=19, where the S/N in our images is high, and where we have a
good chance of detecting fainter companions, disturbed morphologies,
or other potential signs of galaxy-galaxy interactions.  Postage stamp
images of our \Ks$<$19 EROs are shown in Fig.~\ref{morph.fig}.

There are 24 EROs with \RKA\ and \Ks$<$19 in our sample and three of
us (SvdB, MS, and MS) have visually classified their morphologies in
both the $R$ and \Ks\ images simultaneously.  Table~2
compiles the results of this morphological classification.  Of the 24
EROs, eleven show clear evidence for the presence of disks (usually
together with bulges), and five objects have signs of ongoing
interactions or very close companions.

We next ask whether the morphological mix in our ERO overdensity is
different than the mix in the field ERO population.  In doing so we
will be conservative by counting only those objects with clearly
detected disks to give us a lower limit on the disk fraction, and
counting {\it all} those objects without detected disks or signs of
interaction to give us an upper limit on the fraction of passively
evolving ellipticals.  Using this approach we find that {\it at least}
11/24 (i.e., $>$46\%) of the EROs in our field have disks and {\it at
least} 5/24 ($>$21\%) are interacting.  At most 9/24 ($<$38\%) may be
passively evolving ellipticals.  The actual fraction of disks may be
higher than 46\% because faint disks are hard to see against the sky
background; at the same time, the actual fraction of passively
evolving ellipticals may be lower than 38\% because here we have
conservatively assumed that the three point-source objects which are
unresolved are pure-bulge galaxies rather than foreground stars and
that objects that are resolved but for which we were unable to make a
morphological decision nevertheless are pure bulge EROs.  Our \RKB\
sample, although consisting of only ten \Ks$<$19, objects contains a
similar morphological mix of {\it at least} 40\% (4/10) disks and {\it
at most} 50\% (5/10) pure bulges.

Comparing morphological fractions between different surveys is
complicated because different surveys use different morphological
classification schemes and work with imaging data of different
quality.  Moreover, different surveys use different color selection
systems (e.g., $R$$-$\Ks, $I-K$) to define EROs, resulting in
sensitivity to different star formation histories (see Yan \& Thompson
2003 for a discussion) and hence, potentially, to different
morphologies.  Additionally, even when the same filter system is used,
different color cuts within that system (e.g., $R$$-$\Ks$>$5.0, \RKA,
\RKB) select populations spanning different redshift ranges or
obscured by different amounts of interstellar dust (see, e.g.,
Fig.~\ref{modelcol.fig}).  Although all these effects limit the value
of comparisons between different surveys, it is nevertheless
interesting to compare the morphological mix in our ERO overdensity
with that found in field ERO surveys in order to look for gross
trends.

The morphological mix of at least 46\% disks, less than 38\% passively
evolving ellipticals, and at least 21\% interacting systems in our
\RKA, \Ks$<$19 sample is consistent with recent {\it field} ERO studies.
Specifically, our morphological mix is consistent with the HST study
of \RKs$>$5.3 EROs in the GOODS-N field by Moustakas et al.\ (2004)
who found 33-44\% of their EROs to be early-type; it is also
consistent with the HST study by Smith et al.\ (2002) of \RK$>$5.3
EROs who found a 65\% fraction of disks and irregulars; with the Yan
\& Thompson HST study of bright \IK$>$4 EROs who found that 30\% are
bulge-dominated, 65\% are disk-dominated, and 17\% are mergers or
interacting systems; with the Gilbank et al.\ (2003) study of \IK$>$4
EROs who found that 35\% have disk components, up to 30\% are
spheroidal, and 15\% are disturbed or irregular; and with the Roche et
al.\ 2002 ground-based study of \RK$>$5.0 EROs who found a 3:2 mixture
of bulge and disk light profiles with $\sim$25\% showing signs of
interaction.  All these recent morphological studies of {\it field}
EROs find that the population consists of a mixture of bulge- and
disk-dominated systems, with the bulge-dominated systems being in the
minority (at $\sim$1/3 of the total) and, moreover, contains a small
but sizeable fraction of $\sim$1/5--1/4 of interacting systems.  Our
morphological fractions are entirely in line with these field ERO
studies, and we conclude that there is no evidence for difference in
ERO morphological mix between the overdense region and the field.

\section{SUMMARY, DISCUSSION, AND CONCLUSIONS}

In this work we used a deep \Ks-band image of a field with existing
extremely deep \R-band imaging to study the properties of EROs in an
overdense region. These deep images allowed us to robustly select both
\RKA\ and \RKB\ EROs down to \Ks=21 (80\% completeness) over most of
the image.  The ERO surface number density in our field is 3--4 times
higher than in the literature, which represents a very significant
excess even given the strong clustering of EROs.  This excess leads us
to us to conclude that our field may contain a physical structure of
EROs --- possibly a filament or a protocluster --- which could have a
{\it physical} ERO density of up to $\sim$5000 times that of the field
ERO population.

We used the excellent depth of our data (\Kslim=21 over most of the
field) to examine the shape of the ERO number counts in our field.  We
found no evidence in the cumulative ERO surface number density that
the ERO population in our ERO overdensity is different from that in
the field: aside the overall increase in normalization, the surface
number density of EROs in our structure is consistent with that of
field EROs to within the uncertainties down to at least \Ks=20.5 and
includes evidence for a break in slope at \Ks=19--20.

We then use the excellent image quality of our data (0.4\arcsec\ in
\Ks, 0.6\arcsec\ in $R$) to classify the morphologies of bright EROs,
\Ks$<$19, where we have sufficient S/N to detect faint features above
the sky background.  We found that the morphological mix of bright
EROs in our structure is similar to that found in morphological
studies of field EROs.  Specifically, we found that of the 24 \Ks$<$19
EROs in our structure, {\it at least} 11 (46\%) have evidence for
disks and {\it at most} nine (38\%) can be passively evolving
ellipticals.  We also find that at least five (21\%) show signs of
galaxy-galaxy interactions.  These morphological mixes are in line
with morphological fractions of field EROs, and we concluded that
there is no evidence that the cluster and ERO populations differ in
their morphological composition.

The consistency of morphological fractions between our sample and
field ERO samples and the similarity of the shape of the number
counts, both suggest that the ERO population does not vary strongly
with environment.  This lack of environmental differences between
field and overdensity EROs is surprising.  The hierarchical structure
formation paradigm suggests that galaxy formation begins earlier in
overdense regions and consequently we might have expected the ERO
population in the overdensity to be more evolved, with more passively
evolving ellipticals and fewer star-forming disks than in field ERO
samples.  Similarly, if the galaxy formation clock has started at an
earlier redshift in our ERO structure, we might have expected a
steeper faint-end slope of the ERO cumulative surface number density
as suggested by some PLE models.  At present, however, we are forced
to conclude that there is no evidence for environmental dependence of
the ERO galaxy population.

One explanation for this surprising similarity between field and
overdensity EROs is that the EROs in our overdensity did {\it not}
form in it but formed in the field instead (and thus are coeval with
the field ERO population) and fell into the overdensity at a later
time.  The fact that the morphological mix is not significantly
different from the field population and includes a large proportion of
presumably star-forming disks suggests further that the infall was
recent or that the overdense environment is not yet efficient at
modifying --- or has not yed had time to modify --- the morphologies
of the galaxies it contains through gas stripping and other processes
that have been suggested to operate at lower redshifts.

Environmental differences can teach us much about how massive galaxies
as traced by EROs form and evolve at high redshift, and we feel that
the present work is just the beginning of such studies.  Spectroscopic
observations of EROs in our overdense field should prove very valuable
as they would not only allow us to compare spectral characteristics of
our cluster ERO population with field EROs (as studied by, e.g.,
Cimatti et al.\ 2002 and Yan et al.\ 2004a), but would also let us
understand the ERO redshift distribution in our field, potentially
giving important information about the line-of-sight size of the
structure and/or its mass.  Spitzer Space Telescope mid-IR
observations can be used to constrain the number of actively
star-forming galaxies among our cluster EROs, and the comparison of
their abundance with that amongst field EROs (Yan et al.\ 2004b) may
teach us about how star formation in EROs depends on their
environment.  Additionaly, since our field contains a relatively large
number of bright stars (see Fig.~\ref{exptime.fig}), our morphological
study can be improved with adaptive optics imaging.  Finally, other
known ERO overdensities should be {\it systematically} studied to
assemble a larger set from which to draw more robust conclusions.
Several significant ERO overdensities have now been reported in the
literature (e.g., Hall \& Green 1998; Cimatti et al.\ 2000; Best et
al.\ 2003; Toft et al.\ 2003; Wold et al.\ 2003, V\"ais\"anen \&
Johansson 2004b) and deep optical/NIR imaging, mid-IR photometry, and
spectroscopic follow-up will give a valuable data set with which to
study environmental differences that impact the formation and
evolution of massive galaxies.


\vspace{10mm}

We thank the ESO Paranal staff for their work in obtaining these data.
We also thank Jerzy Sawicki, David Thompson, Wayne Barkhouse, and the
anonymous referee for useful comments and discussions.  This work made
use of the NASA Extragalactic Database (NED).  GMO is supported by a
Clay Fellowship at the Smithsonian Astrophysical Observatory.  This
research was partly financed by the Centro de Astrof\'{\i}sica FONDAP,
CONICYT, Chile.



\clearpage

\clearpage

\LongTables

\begin{deluxetable}{lllccc}
\tablecolumns{6}
\tablewidth{5.0truein} 
\tablecaption{ERO Catalog}
\label{catalog.tab}
\tablehead{
\colhead{ID} & 
\colhead{R.A.(2000)\tablenotemark{a}} & 
\colhead{Decl.(2000)\tablenotemark{a}} & 
\colhead{\Ks} & 
\colhead{\RKs} & 
\colhead{Class\tablenotemark{b}}
}
\startdata
 100 & 19:24:04.153 & $-$21:00:16.291 & 19.80$\pm$0.14 & 5.49$\pm$0.18 & R   \\
 107 & 19:24:06.006 & $-$21:00:13.966 & 20.89$\pm$0.24 & 7.75$\pm$1.14 & R   \\
 118 & 19:24:06.639 & $-$21:00:14.985 & 17.52$\pm$0.04 & 6.13$\pm$0.06 & R   \\
 133 & 19:24:16.166 & $-$21:00:13.325 & 18.49$\pm$0.07 & 6.48$\pm$0.10 & R   \\
 178 & 19:24:10.310 & $-$21:00:08.392 & 19.72$\pm$0.14 & 6.56$\pm$0.30 & R   \\
 199 & 19:24:10.014 & $-$21:00:07.581 & 17.86$\pm$0.05 & 5.39$\pm$0.08 & R   \\
 216 & 19:24:16.590 & $-$21:00:01.166 & 16.50$\pm$0.03 & 5.39$\pm$0.04 & R   \\
 262 & 19:24:16.961 & $-$21:00:01.876 & 18.29$\pm$0.07 & 5.70$\pm$0.11 & R   \\
 283 & 19:24:09.002 & $-$20:59:56.030 & 20.76$\pm$0.23 & 6.35$\pm$0.44 & R   \\
 344 & 19:24:13.313 & $-$20:59:51.544 & 20.73$\pm$0.21 & 6.88$\pm$0.47 & U   \\
 403 & 19:24:15.137 & $-$20:59:45.207 & 17.81$\pm$0.05 & 5.37$\pm$0.07 & R   \\
 405 & 19:24:22.275 & $-$20:59:46.121 & 18.99$\pm$0.08 & 7.32$\pm$0.16 & U   \\
 434 & 19:24:04.166 & $-$20:59:44.781 & 19.25$\pm$0.11 & 6.55$\pm$0.21 & R   \\
 500 & 19:24:20.954 & $-$20:59:37.389 & 20.49$\pm$0.20 & 5.33$\pm$0.24 & U   \\
 503 & 19:24:16.621 & $-$20:59:38.249 & 18.97$\pm$0.09 & 5.83$\pm$0.10 & R   \\
 535 & 19:24:17.240 & $-$20:59:33.670 & 18.47$\pm$0.07 & 5.79$\pm$0.11 & R   \\
 553 & 19:24:05.754 & $-$20:59:32.043 & 20.22$\pm$0.16 & 6.02$\pm$0.20 & U   \\
 576 & 19:24:21.612 & $-$20:59:29.631 & 20.77$\pm$0.21 & 6.72$\pm$0.44 & U   \\
 609 & 19:24:21.470 & $-$20:59:28.234 & 19.79$\pm$0.13 & 5.91$\pm$0.19 & U   \\
 612 & 19:24:19.627 & $-$20:59:27.080 & 19.20$\pm$0.09 & 5.36$\pm$0.10 & U   \\
 616 & 19:24:17.366 & $-$20:59:27.731 & 20.33$\pm$0.23 & 5.54$\pm$0.32 & R   \\
 642 & 19:24:21.559 & $-$20:59:24.399 & 20.35$\pm$0.17 & 6.85$\pm$0.34 & U   \\
 674 & 19:24:17.801 & $-$20:59:20.601 & 20.83$\pm$0.22 & 5.86$\pm$0.29 & U   \\
 703 & 19:24:04.883 & $-$20:59:19.663 & 20.27$\pm$0.17 & 5.59$\pm$0.25 & R   \\
 727 & 19:24:19.087 & $-$20:59:17.119 & 19.38$\pm$0.10 & 5.94$\pm$0.13 & R   \\
 733 & 19:24:21.244 & $-$20:59:15.076 & 20.50$\pm$0.19 & 6.54$\pm$0.35 & U   \\
 734 & 19:24:06.165 & $-$20:59:15.414 & 20.20$\pm$0.16 & 6.62$\pm$0.29 & U   \\
 770 & 19:24:17.346 & $-$20:59:11.068 & 18.02$\pm$0.06 & 5.84$\pm$0.08 & R   \\
 812 & 19:24:09.283 & $-$20:59:10.339 & 18.74$\pm$0.08 & 5.46$\pm$0.10 & R   \\
 922 & 19:24:13.296 & $-$20:58:59.767 & 18.91$\pm$0.09 & 5.63$\pm$0.12 & R   \\
 925 & 19:24:17.886 & $-$20:58:57.131 & 20.23$\pm$0.16 & 5.71$\pm$0.25 & R   \\
 958 & 19:24:15.883 & $-$20:58:55.029 & 19.13$\pm$0.10 & 6.10$\pm$0.15 & R   \\
1000 & 19:24:21.840 & $-$20:58:48.904 & 20.53$\pm$0.18 & 5.43$\pm$0.21 & U  \\
1024 & 19:24:03.510 & $-$20:58:47.091 & 19.73$\pm$0.13 & 6.02$\pm$0.16 & U  \\
1041 & 19:24:15.019 & $-$20:58:44.183 & 20.55$\pm$0.21 & 5.39$\pm$0.26 & R  \\
1090 & 19:24:15.735 & $-$20:57:12.278 & 18.80$\pm$0.08 & 5.51$\pm$0.10 & R  \\
1117 & 19:24:20.440 & $-$20:57:14.743 & 20.12$\pm$0.16 & 7.50$\pm$0.60 & U  \\
1141 & 19:24:15.567 & $-$20:57:16.962 & 18.87$\pm$0.09 & 5.58$\pm$0.11 & R  \\
1194 & 19:24:20.899 & $-$20:57:23.016 & 19.95$\pm$0.14 & 6.64$\pm$0.25 & R  \\
1201 & 19:24:10.625 & $-$20:57:24.242 & 18.96$\pm$0.09 & 6.05$\pm$0.12 & R  \\
1205 & 19:24:22.583 & $-$20:57:24.495 & 19.86$\pm$0.14 & 6.77$\pm$0.25 & U  \\
1210 & 19:24:13.322 & $-$20:57:25.375 & 20.14$\pm$0.17 & 5.80$\pm$0.27 & R  \\
1230 & 19:24:10.732 & $-$20:57:28.299 & 20.76$\pm$0.22 & 5.57$\pm$0.27 & U  \\
1260 & 19:24:14.692 & $-$20:57:31.466 & 20.41$\pm$0.17 & 6.86$\pm$0.42 & R  \\
1272 & 19:24:17.349 & $-$20:57:33.207 & 19.58$\pm$0.11 & 5.53$\pm$0.13 & U  \\
1375 & 19:24:17.493 & $-$20:57:43.703 & 19.92$\pm$0.13 & 7.11$\pm$0.28 & U  \\
1448 & 19:24:11.548 & $-$20:57:51.816 & 20.92$\pm$0.22 & 6.28$\pm$0.36 & U  \\
1460 & 19:24:09.808 & $-$20:57:53.466 & 18.80$\pm$0.08 & 5.68$\pm$0.10 & R  \\
1569 & 19:24:15.035 & $-$20:58:02.903 & 20.11$\pm$0.19 & 5.87$\pm$0.31 & R  \\
1646 & 19:24:14.623 & $-$20:58:09.749 & 20.77$\pm$0.24 & 6.58$\pm$0.60 & U  \\
1734 & 19:24:20.134 & $-$20:58:19.264 & 20.33$\pm$0.20 & 5.88$\pm$0.35 & R  \\
1745 & 19:24:15.717 & $-$20:58:19.985 & 18.64$\pm$0.07 & 6.06$\pm$0.10 & R  \\
1789 & 19:24:15.406 & $-$20:58:25.431 & 20.83$\pm$0.21 & 9.12$\pm$3.57 & U  \\
1790 & 19:24:16.733 & $-$20:58:25.371 & 20.21$\pm$0.15 & 5.75$\pm$0.19 & U  \\
1800 & 19:24:15.027 & $-$20:58:26.166 & 20.21$\pm$0.16 & 6.65$\pm$0.29 & U  \\
1812 & 19:24:11.958 & $-$20:58:28.127 & 19.84$\pm$0.15 & 5.70$\pm$0.23 & R  \\
1857 & 19:24:08.225 & $-$20:58:31.597 & 19.82$\pm$0.13 & 5.57$\pm$0.15 & U  \\
1885 & 19:24:07.314 & $-$20:58:34.855 & 20.70$\pm$0.21 & 6.17$\pm$0.35 & U  \\
1890 & 19:24:20.748 & $-$20:58:35.387 & 19.11$\pm$0.09 & 5.44$\pm$0.10 & U  \\
1903 & 19:24:03.393 & $-$20:58:37.094 & 19.26$\pm$0.10 & 5.93$\pm$0.14 & R  \\
1909 & 19:24:05.028 & $-$20:58:37.544 & 20.06$\pm$0.16 & 5.95$\pm$0.25 & R  \\
1910 & 19:24:14.965 & $-$20:58:37.180 & 19.91$\pm$0.13 & 5.56$\pm$0.15 & U  \\
1954 & 19:24:21.713 & $-$20:58:41.015 & 19.59$\pm$0.11 & 6.64$\pm$0.19 & U  \\
1968 & 19:24:06.793 & $-$20:58:41.701 & 20.30$\pm$0.16 & 6.01$\pm$0.24 & R  \\
1975 & 19:24:17.631 & $-$20:58:42.206 & 20.05$\pm$0.15 & 5.51$\pm$0.20 & U  \\
1997 & 19:24:06.774 & $-$20:58:44.191 & 19.91$\pm$0.13 & 6.55$\pm$0.19 & U  \\
2074 & 19:24:11.190 & $-$20:55:34.000 & 20.46$\pm$0.19 & 6.00$\pm$0.30 & U  \\
2077 & 19:24:04.688 & $-$20:55:35.005 & 18.34$\pm$0.07 & 6.13$\pm$0.13 & R  \\
2079 & 19:24:06.088 & $-$20:55:33.537 & 19.98$\pm$0.14 & 5.45$\pm$0.17 & U  \\
2094 & 19:24:07.354 & $-$20:55:35.030 & 20.25$\pm$0.18 & 5.43$\pm$0.22 & R  \\
2147 & 19:24:20.278 & $-$20:55:39.935 & 20.39$\pm$0.18 & 6.91$\pm$0.40 & R  \\
2157 & 19:24:07.113 & $-$20:55:42.978 & 19.96$\pm$0.14 & 5.78$\pm$0.17 & U  \\
2158 & 19:24:11.087 & $-$20:55:40.270 & 20.19$\pm$0.18 & 5.87$\pm$0.26 & U  \\
2169 & 19:24:16.784 & $-$20:55:44.154 & 19.90$\pm$0.15 & 6.74$\pm$0.38 & R  \\
2205 & 19:24:06.131 & $-$20:55:46.907 & 19.22$\pm$0.10 & 6.23$\pm$0.15 & R  \\
2220 & 19:24:18.437 & $-$20:55:49.874 & 19.82$\pm$0.13 & 6.40$\pm$0.24 & R  \\
2222 & 19:24:18.497 & $-$20:55:48.347 & 19.02$\pm$0.10 & 6.65$\pm$0.25 & R  \\
2242 & 19:24:16.721 & $-$20:55:49.549 & 20.19$\pm$0.16 & 6.03$\pm$0.22 & U  \\
2254 & 19:24:22.603 & $-$20:55:44.150 & 20.21$\pm$0.16 & 5.83$\pm$0.21 & R  \\
2273 & 19:24:20.242 & $-$20:55:52.398 & 19.12$\pm$0.09 & 5.32$\pm$0.11 & U  \\
2319 & 19:24:22.274 & $-$20:55:55.092 & 19.28$\pm$0.10 & 5.62$\pm$0.11 & U  \\
2357 & 19:24:19.035 & $-$20:56:00.043 & 17.73$\pm$0.05 & 7.03$\pm$0.06 & U  \\
2358 & 19:24:19.395 & $-$20:55:59.749 & 18.27$\pm$0.06 & 6.05$\pm$0.10 & R  \\
2410 & 19:24:08.136 & $-$20:56:07.100 & 20.21$\pm$0.16 & 6.15$\pm$0.22 & R  \\
2421 & 19:24:18.527 & $-$20:56:07.702 & 18.82$\pm$0.08 & 6.10$\pm$0.10 & R  \\
2475 & 19:24:09.125 & $-$20:56:11.069 & 20.79$\pm$0.21 & 5.94$\pm$0.27 & U  \\
2482 & 19:24:19.322 & $-$20:56:15.024 & 19.29$\pm$0.10 & 6.63$\pm$0.16 & R  \\
2509 & 19:24:19.068 & $-$20:56:15.785 & 19.22$\pm$0.10 & 5.82$\pm$0.12 & U  \\
2588 & 19:24:21.415 & $-$20:56:14.121 & 20.30$\pm$0.17 & 6.05$\pm$0.23 & R  \\
2628 & 19:24:11.911 & $-$20:56:27.323 & 20.79$\pm$0.21 & 7.99$\pm$1.27 & U  \\
2659 & 19:24:05.032 & $-$20:56:32.229 & 19.42$\pm$0.11 & 8.14$\pm$0.84 & R  \\
2742 & 19:24:10.715 & $-$20:56:40.164 & 20.30$\pm$0.16 & 5.35$\pm$0.19 & R  \\
2756 & 19:24:20.964 & $-$20:56:40.937 & 20.11$\pm$0.18 & 5.82$\pm$0.30 & R  \\
2791 & 19:24:19.299 & $-$20:56:43.208 & 19.77$\pm$0.13 & 6.59$\pm$0.24 & R  \\
2793 & 19:24:21.152 & $-$20:56:44.473 & 17.85$\pm$0.05 & 6.29$\pm$0.09 & R  \\
2818 & 19:24:20.345 & $-$20:56:45.692 & 18.40$\pm$0.07 & 5.52$\pm$0.09 & R  \\
2899 & 19:24:04.041 & $-$20:56:54.393 & 20.99$\pm$0.27 & 6.52$\pm$0.59 & U  \\
2947 & 19:24:14.613 & $-$20:56:58.961 & 18.56$\pm$0.07 & 5.82$\pm$0.11 & U  \\
2985 & 19:24:15.132 & $-$20:57:02.388 & 19.27$\pm$0.10 & 6.03$\pm$0.12 & U  \\
3012 & 19:24:15.093 & $-$20:57:06.766 & 19.77$\pm$0.13 & 5.37$\pm$0.15 & R  \\
3033 & 19:24:22.498 & $-$20:57:07.613 & 19.35$\pm$0.11 & 5.67$\pm$0.14 & R  \\
\enddata
\tablenotetext{a}{Positions taken from the World Coordinate System of the 
\R-band image. }
\tablenotetext{b}{Star-galaxy separation: R --- Resolved, U --- Uncertain}
\end{deluxetable}

\begin{deluxetable}{llll}
\tablewidth{0pt} 
\tablecaption{\label{morph.tab}Morphological Properties of \Ks$<$19 EROs}
\tablehead{
\colhead{ID} & 
\colhead{\Ks} & 
\colhead{\RKs} & 
\colhead{morphology\tablenotemark{a}}
}
\startdata
 118 & 17.52 &	6.13 & BD	\\
 133 & 18.49 &	6.48 & BD	\\
 199 & 17.86 & 	5.39 & BD	\\
 216 & 16.50 &	5.39 & ?	\\
 262 & 18.29 &	5.70 & I	\\
 403 & 17.81 &	5.37 & I	\\
 405 & 18.99 &	7.32 & U	\\
 503 & 18.97 &	5.83 & ?	\\
 535 & 18.47 &	5.79 & DI	\\
 770 & 18.02 &	5.84 & BD	\\
 812 & 18.74 &	5.46 & BD	\\
 922 & 18.91 &	5.63 & I	\\
1090 & 18.80 &	5.51 & BD	\\
1141 & 18.87 &	5.58 & BD	\\
1201 & 18.96 &	6.05 & ?	\\
1460 & 18.80 &	5.68 & ?	\\
1745 & 18.64 &	6.06 & ?	\\
2077 & 18.34 &	6.13 & BD	\\
2357 & 17.73 &	7.03 & U	\\
2358 & 18.27 &	6.05 & I	\\
2421 & 18.82 &	6.10 & ?	\\
2793 & 17.85 &	6.29 & BD	\\
2818 & 18.40 &	5.52 & BD	\\
2947 & 18.56 &	5.82 & U	\\
\enddata
\tablenotetext{a}{D: disk, B: bulge, I: signs of interaction or close companion, U: morphologically uncertain (possible star --- see \S~\ref{stargalaxy}), ?: galaxy but otherwise morphologically unclassified (i.e., not a star)}
\end{deluxetable}


\end{document}